\documentclass[showpacs,twocolumn,aps]{revtex4}
\usepackage[dvips]{graphicx}
\usepackage{mathrsfs}
\usepackage[dvips,usenames]{color}

\begin{document}
\title{Shannon capacity of nonlinear regenerative channels}
\author{M. A. Sorokina} 
\affiliation{Aston Institute of Photonic Technologies, Aston University, B4 7ET Birmingham, UK}

\author{S. K. Turitsyn}
  \affiliation{Aston Institute of Photonic Technologies, Aston University, B4 7ET Birmingham, UK}

\begin{abstract}
We compute Shannon capacity of nonlinear channels with regenerative elements.  Conditions are found under which capacity of such nonlinear channels is higher than the Shannon capacity of the classical linear additive white Gaussian noise channel. We develop a general scheme for designing the proposed channels and apply it to the particular nonlinear  sine-mapping. The upper bound for regeneration efficiency is found and the asymptotic behavior of the capacity in the saturation regime is derived.
\end{abstract}

\maketitle

Information theory provides underlying concepts for major areas of
science and technology from communications and computer science to
biology and economics. The seminal Shannon's result for capacity
of the linear additive white Gaussian noise (AWGN) channel: $C= B
\, \log_{2} (1 +S/N)$ \cite{Shannon} is arguably one of
the few most known equations in the history of science.  In
practical communication channels that are close to the
linear AWGN channel the Shannon capacity can be approached very
closely by using low density parity check codes and turbo codes.
The linear AGWN channel is nowadays  a textbook material and is
virtually  wholly understood. There is, however, a growing
interest in studies of a capacity of more complex communication
channels including nonlinear channels for which the limits have
yet to be defined. A well known and extremely important for
practical applications example
 is the optical fibre channel that is inherently nonlinear due to intensity-dependent refractive index in silica at high enough signal powers and very small fibre core where signal is guided over long distances.

The definition of the  Shannon capacity for arbitrary channel (in what follows, capacity C is per unit bandwidth) involves maximizing the mutual information functional \cite{Shannon}:
\begin{equation} \label{C_def}
C = \max_{P(x)} \int \mathrm{D}  x \mathrm{D}  y  P(x) P(y|x)
\log_2\frac{P(y|x)}{\int \mathrm{D} x P(x) P(y|x)},
\end{equation}
over all valid input probability distributions $P(x)$ subject to
power constraint $ \int \mathrm{D} x \,P(x) \, |x|^2 \leq S$. Here
statistical properties of the channel are given by the conditional
input-output probability density function (PDF) $P(y|x)$. Thus,
from the view point of the information theory there is not much
difference between linear or nonlinear channels as long as PDF
$P(y|x)$ is known. Yet, the modern information theory is mostly
developed for linear channels, simply because it is often
technically challenging to derive $P(y|x)$ for practical nonlinear
channels. This reflects both difficulty of analysis of nonlinear
systems with noise and the fact that there are varieties of
nonlinear communication channels that hardly can be described by
any single generic theory. An important new feature introduced by
nonlinearity is the possibility of  nonlinear filtering and/or
signal regeneration.  Whenever the nonlinear transformation has
multiple fixed points, the consequent interleaving of the
accumulating noise with nonlinear filter will produce effective
suppression of the noise. In other words the nonlinear filter will
attract a signal to the closest fixed points and suppress the
effective signal diffusion caused by the noise.  Similar idea has
been discussed in multiple contexts from physical systems to the
interpretation of biological memory effects in terms of potentials
with multiple minima \cite{bio1}. In quantum theory a
qualitatively similar phenomena is known under the name of the
Zeno effect where continuous measurement of the quantum system and
associated von Neumann collapse of wave function prevents the
natural dispersion of the wave function and causes the quantum
system to remain in the same state \cite{Zeno1,Zeno2,Zeno3,Zeno4}.
Capacity of nonlinear transmission channels with "hard decision" regenerators have
been examined in \cite{OLTT}.

In this Letter we introduce new method -- \textit{regenerative
mapping} for designing nonlinear information channels with
capacity exceeding Shannon capacity for linear AWGN channel. We
start from substantial expansion of analysis presented in
\cite{OLTT} and comprehensively quantify improvement in the Shannon capacity that
 various nonlinear channels with regenerators can provide over the linear AWGN
channel. Next we introduce new type of channels with smooth nonlinear
transfer functions that are principally distinct from a "hard
decision" regeneration considered in \cite{OLTT}. We stress that
the proposed channel model is fundamentally different from
Decode-and-Forward channel model and does not assume any
decoder/encoder pair to be used with  in-line elements. The
proposed regenerative element acts like \emph{nonlinear filter} on
the stochastic signal distortions through an effective periodic
potential creating attraction regions in the signal mapping.

\begin{figure}
\begin{center}
\includegraphics[width=8cm]{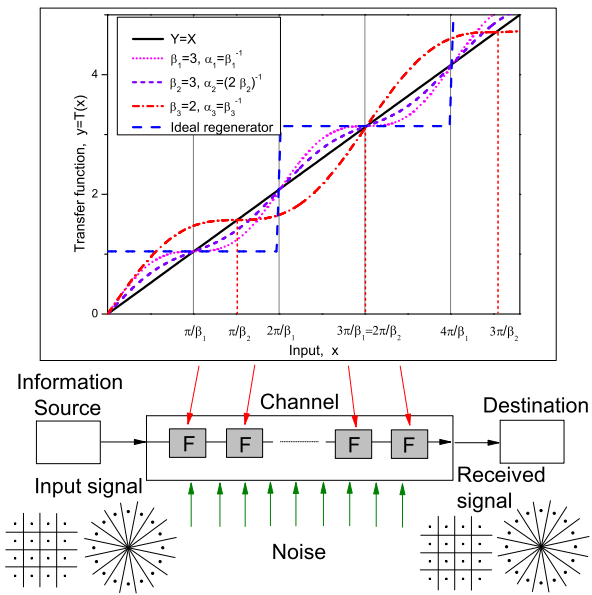}
\end{center}
\caption{  \label{Fig:Scheme} (Color online) The regenerative channel scheme: nonlinear filters are placed equidistantly along the transmission line. Stochastic distortion in the channel is modelled by AWGN. The nonlinear filtering transfer function (TF) is plotted for
the ideal regenerator (long-dashed line) and for the continuous mapping $ y=x+ \alpha \sin (\beta x)$ with parameters: (a) $\beta_{1,2}=3$, superstable
$\alpha_1=1/\beta_1$ (dotted curve) and stable $\alpha_2=1/(2\beta_2)$ (short-dashed
curve) points; (b) $\beta_3=2$ and superstable point $\alpha_3=1/\beta_3$ (dashed-dotted curve). The alphabet is
shown by vertical grey solid ($\beta_3=3$) and red dotted ($\beta_3=2$) lines. Voronoi diagrams (decision boundaries) for ideal regenerators, i.e. attraction regions created by nonlinear TF are shown for 16 - rectangular and ring packing. }
\end{figure}

Consider the nonlinear regenerative channel model with $R$
identical nonlinear filters placed in the transmission line. The
signal transmission is distorted by stochastic process which is
modeled as AWGN uniformly distributed
along the line. 
%
%
%
The  noise term incorporates the stochastic effects from different
sources. Depending on the model application, the stochastic
process can be considered as an analogue of the random force in
the time-continuous case or noise that adds up to signal
transmission through the media. This term also includes  an
additive noise that is originated from by the nonlinear device
itself.  Here we develop
the general mathematical method of constructing and optimizing the
set of transfer functions $y=T(x)$ (see e.g. Fig.
\ref{Fig:Scheme}) that can increase nonlinear transmission system
capacity beyond the capacity of the linear AWGN channel.

Shannon capacity (Eq. \ref{C_def}) of the considered  systems is a
function of signal to noise ratio (SNR), the number of nonlinear filters $R$ and the
parameters of nonlinear mapping. SNR is defined here as the ratio of
the input signal power $S$ to the noise added linearly to the
signal during transmission at each node $k=1,...,R$:
 \[SNR=\frac{S}{N},  \ \ \ N=\sum_{k=0}^{R} N_k\]

 Note that in the nonlinear communication system due to the mixing of signal with noise during propagation the definition of in-line SNR is a nontrivial issue. The introduced SNR has the meaning of the signal to noise ratio in the respective linear system in the absence of nonlinear in-line elements. This allows us to compare performance of the considered system with the corresponding  linear AWGN channel having the same noise level. Evidently, that the effect of noise squeezing is enhanced with the number of regenerators/nonlinear filters. Therefore, to evaluate the system performance improvement one need to take into account the accumulative effect of nonlinear signal and noise transformation along the line. To quantify the overall effect, we need to study capacity of the source-destination transmission as a function of signal power ratio to the sum of power of all added noise  at the source-destination link. 

 We start the capacity analysis with the system of the ideal regenerators that is used further as a reference system. In this case, we can derive analytical expressions for the maximum achievable regenerative capacity gain. Next, we introduce the general method of designing and optimization for the class of regenerative channels that guarantee capacity improvement without requirement of decoder/encoder pair. Finally, we illustrate the proposed method by presenting new nonlinear channels with capacity above the Shannon capacity of the linear AWGN channel.

A physical model of regenerative mapping  for increasing information capacity can find applications in various fields. The considered  model can be
applied to the evolution of a stochastic system in time with
$k=0,1,...,R$ being a temporal index (see \cite{math1},
\cite{math2}) or to analysis of a spatial dynamics, when signal
propagates through the cascaded scheme of regenerative filters.
The latter model, for instance, can be applied to signal
propagation and regeneration in fiber-optic communication (see
e.g. \cite{OR1}-\cite{Slavik} and references therein). The ring
packing regeneration was recently demonstrated by phase sensitive
amplification \cite{Slavik}. We would like to emphasise in this
Letter generic features of the model, important for both spatial
and temporal applications.

We start the analysis with the ideal regenerators that assign each
transmitted symbol to the closest element of the given alphabet
(in Fig. \ref{Fig:Scheme}, the corresponding stepwise transfer
function is plotted by the dashed blue line). The step function defines a
maximum regeneration capability and, consequently, determines the
maximum achievable capacity gain due to nonlinear filtering. The
conditional pdf of such a system  is defined through matrix
elements \cite{OLTT}:
\begin{equation} \label{Pid}
P(y=x_k|x_l)=\int_{S_k} dx' P_C(x'|x_l)=W_{kl} ,
\end{equation}
where transition matrix is defined as follows
\[
W_{kl}=\frac{1}{2}\Big(\mathrm{erf}[\Delta_{kl}^+]-\mathrm{erf}[\Delta_{kl}^-]\Big),
\Delta_{kl}^\pm=(x_k+x_{k\pm1}-2x_l)\sqrt{\frac{R}{8N}},
\]
Due to Markovian nature of the stochastic system, the overall
transition matrix after $R$ regenerative segments reads as $W^R$.


First, consider the \emph{rectangular constellations} (also known as quadrature amplitude modulation).
At low SNR range the channel is
binary in each of $n$ dimension. Therefore, capacity is well approximated by the following expression:
\begin{equation} \label{lCr}
C_R\simeq\underline{C_R}= n(1+m_{+}log_2(m_{+})+m_\pm\log_2m_\pm)
\end{equation}

with the transition matrix elements(denote SNR as $\rho$):
\[m_{+}=m_{-}=\frac{1}{2^R}\Big(1+\mathrm{erf}\Big[\sqrt{\frac{R\rho}{2}}\Big]\Big)^R, \ \
m_\pm=1-m_{+}\]

As SNR increases, the closest neighbors distance reaches the optimal cell size which is defined by the noise variance and the
number of in-line regenerators, $d^2_{opt}=16N/R \ \Omega[e^2 R^2/(16 \pi)]$, here  $\Omega$ is the so-called Lambert function.
  Thus, at high SNR the system is characterized by the optimal decision boundaries that are
sufficiently large compared with the noise variance to suppress
noise effectively. Therefore, with the growing signal power the
amplitude distribution $x_l$ remains constant (i.e. equidistant
with the closest neighbors distance $d_{opt}$), whereas the
maximum entropy principle defines Maxwell-Boltzmann distribution
as the optimal pdf for a fixed average energy constraint.
Thereafter, the output pdf can be well approximated as $q_l=\nu
e^{-\lambda x_l^2}$, here constants are chosen to satisfy
conditions $\sum_{l=1}^M q_l=1$ and $\sum_{l=1}^M q_l x_l^2=S+N/R$
($\lambda =1/[2(S+N/R)]$ for the large number of constellation
symbols). In the limit of high SNR and/or large number of
regenerators, so that $\Delta=\sqrt{2\Omega[e^2 R^2/(16 \pi)]}\gg
1$, the noise is sufficiently squeezed and the faulty decision
occurs only between the nearest neighbors. The symmetry of the
modulation format allows us to consider one-dimensional $M$ points
problem, the resulted capacity is multiplied by the number of
dimensions $n$. Finally, the channel capacity in the limit of high
SNR for M-rectangular discrete modulation format is found as
\begin{equation} \label{hCr}
\overline{C_R} = n\sum_{l=1}^M q_l\log_2 q_l+nR\frac{e^{-\Delta^2}}{\Delta\sqrt{\pi}}\log_2\Big[R\frac{e^{-\Delta^2}}{4\Delta\sqrt{\pi}}\Big]
\end{equation}

Hence, with growing SNR one can observe a constant gap (that
quantifies improvement) between the regenerative channel and
linear AWGN channel capacities. The capacity improvement factor is
defined by the noise variance and the number of regenerators:
\begin{equation} \label{dCr}
\Delta C_R=\frac{n}{2}\log_2\Big(\frac{\pi
eR}{4\Delta}\Big)+nR\frac{e^{-\Delta^2}}{\Delta\sqrt{\pi}}\log_2\Big[R\frac{e^{-\Delta^2}}{4\Delta\sqrt{\pi}}\Big]
\end{equation}

\begin{figure}
\begin{center}
\includegraphics[width=8cm]{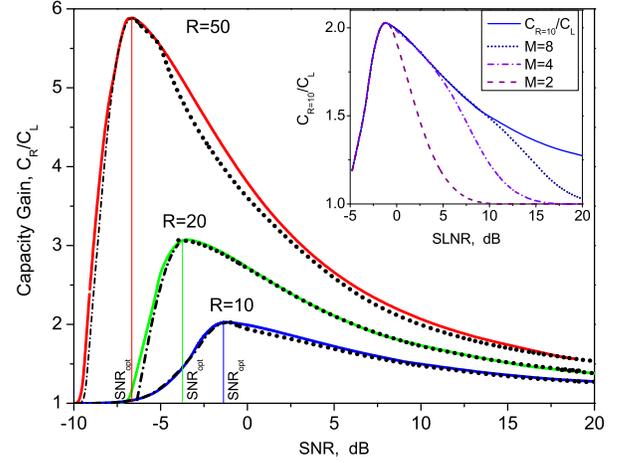}
\end{center}
\caption{  \label{Fig:CCs} (Color online) Gain, the capacity ratio
to the Shannon formula $C_L=\log_2(1+\rho)$, for the different
number of regenerators. The analytical results demonstrating an excellent agreement with numerics are shown by
black (dash-dotted Eq.\ref{lCr} and dotted Eq.\ref{hCr}) lines. The inset shows mutual information gain for
$M^2$- rectangular constellation. }
\end{figure}

\begin{figure}
\begin{center}
\includegraphics[width=8cm]{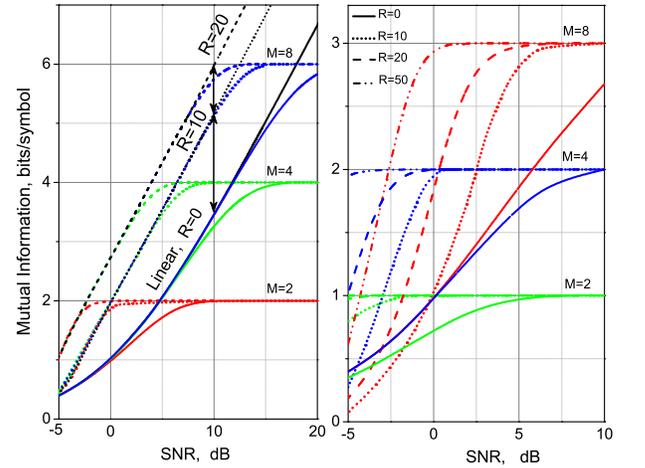}
\end{center}
\caption{  \label{Fig:C_rings} (Color online) Capacity compared to MI for discrete $M^2$ - rectangular (left panel) and M - points in the ring (right panel) alphabets. The arrows show the constant capacity increase. }
\end{figure}

Thus, the maximum  capacity gain due to regeneration
(i.e. the maximum regeneration efficiency) is observed for the
binary channel. This result emphasizes efficiency of a simple binary channel
that might be important for practical design consideration.
 The minimum SNR value,
when ${d_{opt}}$ is achieved, defines the maximum capacity ratio
to its linear analogue,
i.e.\[SNR_{opt}=\frac{d_{opt}^2}{4N}=\frac{4}{R} \
\Omega\Big(\frac{e^2 R^2}{16 \pi}\Big)\] Also, at this SNR value
two analytic formula Eq. \ref{lCr}-\ref{hCr} can be interpolated
to describe capacity at the full range of SNR values.

The analytical approximations shown in Fig. \ref{Fig:CCs} by
black line demonstrate an excellent agreement with the result of numerical computations of the
capacity for  different number of regenerators. The figure
demonstrates that regeneration allows substantially reduce noise
impairments and, consequently, achieve capacity above the Shannon
linear capacity. The maximum system improvement is observed at
low SNR values -- $SNR_{opt}$ shown by vertical lines. Fig.
\ref{Fig:C_rings} shows capacity for rectangular (left panel) and
ring (right panel) packing (in communications \cite{FC2}, those
constellations correspond to the so-called quadrature amplitude
modulation and phase-shift-keying modulation formats,
respectively). Here the role of optimization is demonstrated: the
optimal constellation choice depends on the SNR value and the
number of the in-line regenerators. It is important to note, that
with the non-optimal format one can get the resulting nonlinear
capacity lower than the linear AWGN channel Shannon capacity (due
to non-optimal value of the cell size), whereas the correct format
choice can give dramatic gain in capacity.


Further, to demonstrate the \emph{constructive role of
nonlinearity} without necessarily "hard decision", we propose the nonlinear filtering element that
transforms the input $x$ into the output $y$ in the following way: $y=T(x)=x+\alpha \sin(\beta x)$. The model is commonly referred in theoretical physics as a sine-mapping (see \cite{math1}, \cite{math2}).
One of the possible ways to achieve sine-mapping is to apply sine-Fourier
transformation $\mathbf{F}_s[s(x)]=\int_{-\infty}^\infty dx \sin
(\beta x) s(x)$ to each quadrature $z=(x_R,x_I)$ of the input
signal described by the waveform $s=\sum_l \delta(z-z_l)f(t-lT_s)$
(\cite{FC2}), here summation is performed over the number of symbols and $T_s$ is a symbol period.
%
Then we add up the original and the transformed signal amplified by
$\alpha$. The resulted transformation for each quadrature $z=(x_R,x_I)$ is then given by
\[y=x+\alpha \sin(\beta x)\]

Note that the  sine-mapping nonlinear transfer function is the continuous
analogue (first approximation) of the Fourier series expansion of
TF for the ideal stepwise regenerator.

The signal evolution in such system can be presented by the
stochastic map -- a discrete version of the Langevin equation for
stochastic processes:
\[ y_{k}=T(y_{k-1})+\eta_{k}, \ \ \ k=1,...,R,  \ \ \ y_0=x+\eta_0, \,y=y_{R}\]

Here $k$ is the discrete spatial/temporal index and $T$ is the
transfer function of the regenerative filter (see the channel
scheme in Fig. \ref{Fig:Scheme}). The  term $\eta_k$ models the
Gaussian noise with zero mean and the variance given by $N_k$
added at $k$-th node. The model scheme is shown in Fig.
\ref{Fig:Scheme}. The nonlinear map has a set of special points
that are optimal for nonlinear filtering. The sine-defined transformations
result in the effective periodic potential which creates
attraction regions in the signal mapping without making
the hard decision. Whereas, the points are "attracted" to the
alphabet, the alphabet should remain stable. This results in the
following set of conditions imposed on the transfer function:
\begin{equation} \label{1}
T[x^*]=x^*, \ \ T''[x^*]=0, \ \ |T'[x^*]|<1
\end{equation}
The first expression implies that the alphabet is defined by the
stationary points $x^*$ of the mapping. Next, the transfer function should change
curvature at the alphabet,i.e. the alphabet point is the inflection point of the transfer function. The third expression reflects stability, so that the the signal points distortion is effectively suppressed. When the first derivative is equal to zero, the alphabet is
superstable.  

In the considered example of sine
mapping, the alphabet is placed at the points $\pi(2k+1)/\beta$ where $k\epsilon
\mathrm{Z}$ that are stable if and only if
$\alpha\beta\leq1$. Note that, in particular, the system is superstable when
$\alpha\beta=1$

\begin{figure}
\begin{center}
\includegraphics[width=9cm]{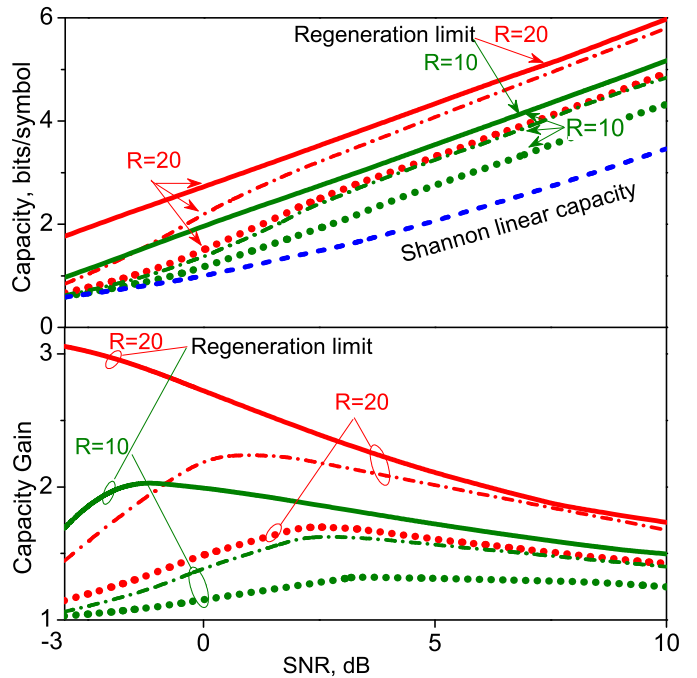}
\end{center}
\caption{  \label{Fig:CSine} (Color online) Numerically calculated
capacity and gain in the capacity (compared to linear AWGN channel) for the  channel with $R=20$ and $R=10$ nonlinear in-line
filters (showed by red and green color correspondingly). The optimization was performed over the filter parameter
$\beta$ (which defines the alphabet) and input pdf. Both the
superstable (dashed-dotted curves) $\alpha\beta=1$ and stable (dotted
curves) $\alpha\beta=0.5$ mapping was considered. The upper bound of regeneration with the given number of in-line filters R=10 and R=20 is showed by solid green and red curves, respecively. }
\end{figure}

The conditional pdf for the output at k-th node for each quadrature $y_k$
given the input $y_{k-1}$ is found as
\[P(y_k|y_{k-1})=\frac{1}{\sqrt{\pi N_{k}}}\exp\Big[-\frac{|y_k-T(y_{k-1})|^2}{N_{k}}\Big]\]
Because of the Markovian property of the process, the conditional
pdf of the received signal after propagation through $R$ links,
$y_{R}$ given the input $x$ is expressed by a product of
single-step conditional probabilities
\begin{equation} \label{Pyx}
P(y_{R}|x)=\int dy_{R-1}..dy_0 P(y_{R}|y_{R-1})...P(y_0|x)
\end{equation}

Consequently, when $N_k=N_0=\mathrm{const}$ the conditional pdf can be expressed through Onsager-Machlup functional
or action of the
path given by $\mathcal{S}=\Sigma_k (y_k-T(y_{k-1}))^2$ as follows
\[P(y_{R}|x)=\int \prod_{k=0}^{R-1}dy_k e^{-\mathcal{S}/N_0}\]

The numerically calculated capacity is shown in Fig. \ref{Fig:CSine}. The proposed sine-mapping nonlinear channel demonstrates visible capacity gain over the Shannon capacity of
the linear AWGN channel. Here we demonstrate that though TF with plateau is the most efficient, nevertheless, it is not necessary requirement for regeneration.  One can see that suboptimal
parameter values also provide a capacity increase. The set of conditions given bye Eq.\ref{1} defines optimization and design rules for implementation of such nonlinear regenerative channels.

  In the limit of large SNR  and/or large number of nonlinear filters all regenerative schemes tend to the asymptotic behaviour,
 when the gain gap between regenerative and linear AWGN channel capacity is constant. The saturation effect occurs when noise is squeezed to
  such level that the stochastic distortion is small and the shift takes place within the plateau area. Thus, under such conditions
  the system with the nonlinear filter is equivalent to the considered above ideal regenerative system. In the limit of
  small noise the core in Eq.\ref{Pyx} approximates delta-function behavior and, consequently, reproduce the result of Eq. \ref{Pid}.
   Thus, the capacity gain of any optimized regenerative system considered here tends to the asymptotic value defined by Eq.\ref{dCr}. 
     Using the method of steepest descent we derive the capacity gain for the sine transfer function with the sub-optimal parameters relation (see Eq. \ref{1}) $q=\alpha \beta\leq1$:
\[\lim_{SNR\rightarrow\infty}\Delta C_S = \Delta C_R(R)-\frac{n}{2}\log_2\Big(\frac{1-(1-q)^{2(R+1)}}{1-(1-q)^2}\Big)\]

We develop analytical model that proves the information capacity increase in stochastic systems with regenerative mapping. The gain is achieved by the
noise squeezing due to introduced nonlinear filter design that creates attraction regions around the stable alphabet.
We presented the design rules for implementation of nonlinear regenerative systems with increased  channel capacity.
The introduced classes of nonlinear devices can be used for construction of nonlinear communication channels with capacity exceeding the Shannon capacity of the linear
AWGN channel. We anticipate that our results will lead to new insights into the Shannon capacity of nonlinear communications channels.

We are grateful to K. Turitsyn and E. Narimanov for useful discussions.
The support under the UK EPSRC Programme Grant UNLOC EP/J017582/1 is gratefully acknowledged

\end{document}